\documentclass[twocolumn,10pt]{IEEEtran} 
\usepackage[utf8]{inputenc}
\usepackage{epsfig,cite}
\usepackage{graphicx}
\graphicspath{{graphics/}}
\usepackage{color}
\usepackage{multirow} 
\usepackage{makecell} 
\usepackage{amssymb,amsmath}
\usepackage{mathrsfs} 

\newcommand{\edit}[2]{{#1}}

\IEEEoverridecommandlockouts

\DeclareMathOperator{\erf}{erf}

\begin{document}
\bibliographystyle{IEEEtran}

\title{Modeling Interference-Free Neuron Spikes with Optogenetic Stimulation}
\author{
	\IEEEauthorblockN{Adam Noel$^1$, Shayan Monabbati$^2$, Dimitrios Makrakis$^3$, Andrew W. Eckford$^4$}
	\IEEEauthorblockA{$^{1}$School of Engineering, University of Warwick, Coventry, UK, Email: adam.noel@warwick.ac.uk\\
		$^{2,4}$Department of EECS,
		York University,
		Toronto, Ontario, Canada, Email: $^2$shayan.monabbati@case.edu, $^4$aeckford@yorku.ca\\
	$^{3}$School of EECS, University of Ottawa, Ottawa, Ontario, Canada, Email: dimitris@eecs.uottawa.ca}
}

\author{Adam Noel\IEEEauthorrefmark{1}, \IEEEmembership{Member, IEEE}, Shayan Monabbati, \IEEEmembership{Student Member, IEEE}, Dimitrios Makrakis, and Andrew W. Eckford, \IEEEmembership{Senior Member, IEEE}
	\thanks{Manuscript draft. A preliminary version of this work was presented at IEEE ICC 2018 \cite{Noel2018}. This work was supported in part by the Natural Sciences and Engineering Research Council of Canada (NSERC). \emph{Asterisk indicates corresponding author}.}
	\thanks{\IEEEauthorrefmark{1}A.~Noel is with the School of Engineering, University of Warwick, Coventry, UK (email: adam.noel@warwick.ac.uk).}
	\thanks{S.~Monabbati is with the Department of Electrical Engineering and Computer Science, Case Western Reserve University, Cleveland, OH, USA.}
	\thanks{D.~Makrakis is with the School of Electrical Engineering and Computer Science, University of Ottawa, Ottawa, ON, Canada.}
	\thanks{A.~W.~Eckford is with the Department of Electrical Engineering and Computer Science, York University, Toronto, ON, Canada.}
}

\newcommand{\EXP}[1]{\exp\left(#1\right)}
\newcommand{\ERF}[1]{\erf\left(#1\right)}
\newcommand{\powten}[1]{\times 10^{#1}}

\newcommand{\metre}{\textnormal{m}}
\newcommand{\second}{\textnormal{s}}
\newcommand{\mol}{\textnormal{mol}}

\makeatletter
\newcommand{\vast}{\bBigg@{3}}
\newcommand{\Vast}{\bBigg@{4}}
\makeatother

\newcommand{\light}{\ell}
\newcommand{\rateTarget}{\lambda_\textrm{T}}
\newcommand{\fireThresh}{v_\textrm{fire}}
\newcommand{\tMin}{t_\textrm{min}}
\newcommand{\nMin}{{n_\textrm{min}}}
\newcommand{\tSep}{t_\Delta}
\newcommand{\nSep}{n_\Delta}
\newcommand{\distortion}{d}

\newcommand{\Imax}{I_\mathrm{max}}
\newcommand{\toff}{t_\mathrm{off}}
\newcommand{\ton}{t_\mathrm{on}}
\newcommand{\tauon}{\tau_\mathrm{on}}
\newcommand{\tauoff}{\tau_\mathrm{off}}
\newcommand{\dt}{\Delta t}

\maketitle

\begin{abstract}
This paper predicts the ability to externally control the firing times of a cortical neuron whose behavior follows the Izhikevich neuron model. The Izhikevich neuron model provides an efficient and biologically plausible method to track a cortical neuron's membrane potential and its firing times. The external control is a simple optogenetic model represented by \edit{an illumination source that stimulates a saturating or decaying membrane current}{R2C3}. This paper considers firing frequencies that are sufficiently low for the membrane potential to return to its resting potential after it fires. The time required for the neuron to charge and for the neuron to recover to the resting potential are numerically fitted to functions of the Izhikevich neuron model parameters and the \edit{peak}{} input current. Results show that simple functions of the model parameters and \edit{maximum}{} input current can be used to predict the charging and recovery times, even when there are deviations in the actual parameter values. Furthermore, the predictions lead to lower bounds on the firing frequency that can be achieved without significant distortion.
\end{abstract}

\maketitle



\section{Introduction}


Over the past decade, developments in {\em optogenetics} have given researchers the ability to directly stimulate neurons \cite{Fenno2011,Deisseroth2011}.
Using this technique, neurons are modified with a gene that encodes a light-sensitive protein (i.e., an {\em opsin}), causing the neurons to express opsins on their surface. Certain opsins, such as channelrhodopsin \cite{Nagel2002}, open an ion channel in response to light. When the channels are open, an ion current flows through the neuron's membrane, changing its electric potential and causing it to fire. Thus, if an optogenetically-modified neuron is stimulated with a strong light source, such as a laser, then the neuron will eventually fire in response. 

Dramatic advances in the study of the brain, as well as revolutionary new therapies for neurological disorders, are expected to follow from precise optogenetic control over neural circuits \cite{Deisseroth2012}. 
So far, research has often focused on the control of large groups of neurons in experimental settings \cite{Grosenick2015}; e.g., studies of seizures in the mouse brain \cite{Armstrong2013} or of spinal cord injury in rats \cite{Wenger2014}. However, targeted control of individual neural circuits are of considerable interest, and recent experimental results have demonstrated the feasibility of this approach \cite{Andrasfalvy2010,Packer2013,Shemesh2017}. It is widely expected that this control will one day lead to optogenetics-based therapies for neurological problems \cite{Rajasethupathy2016}, such as epilepsy \cite{Tonnesen2009} or recovery from neural injury \cite{Li2011}. \edit{Within the domain of communication and networking, controlled synaptic communication has been proposed as a potential communication technique for nanonetworks \cite{Veletic2016,veletic2019}, including potential interfaces between neurons and nanomachines \cite{Mesiti2013}. Communication models of synaptic systems have also been produced \cite{tagluk2019,malak2013,khan2019}. Furthermore, optogenetic techniques have been discussed in the context of therapeutic nanonetworks \cite{Wirdatmadja2017,Balasubramaniam2018} and for brain-machine interfaces \cite{jornet2018}.}{R1C2}

\begin{figure}[!t]
    \centering
    \includegraphics[width=3.4in]{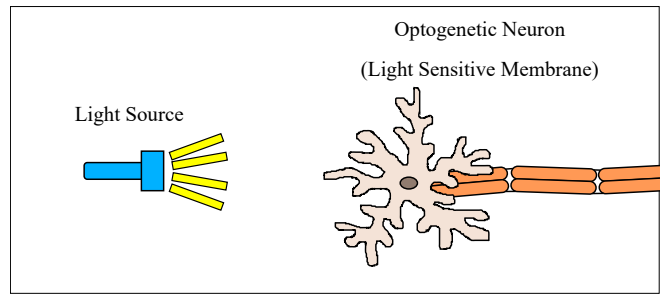}
    \caption{Conceptual diagram of the system model. A neuron with light-sensitive opsins expressed on its surface is stimulated by a light source.}
    \label{fig_system_model}
\end{figure}

\edit{For these applications,}{} an interesting problem is to precisely control the firing time of an individual neuron, as shown conceptually in Fig.~\ref{fig_system_model}. 
Consider a neuron illuminated by a light source, where $\ell(t)$ is the time-varying light intensity. Let $\mathbf{t} = [t_1,t_2,\ldots,t_n]$ represent a vector of times at which the neuron fires. Then the neuron may be viewed as a functional $\mathsf{n}(\cdot)$, taking $\ell(t)$ as input and returning $\mathbf{t}$. The control problem is to invert $\mathsf{n}(\cdot)$: that is, given a desired vector $\mathbf{t}$, find $\ell(t)$ as a solution for $\mathbf{t} = \mathsf{n}(\ell(t))$.
The solution to this problem strongly depends on the neuron model $\mathsf{n}(\cdot)$, for which different models exist.

There have been various approaches to this problem in the recent literature. Some approaches treat the optogenetic stimulus and response as a control system \cite{newman2015,quine-arXiv}, an approach that has led to designs for therapeutic medical devices \cite{mickle2019}. Other approaches have focused on detailed neuron models with optogenetic ion channels (particularly channelrhodopsin) \cite{renault2018}, or models based on photoconversion \cite{olson2017}.

%
%

The simple, yet tractable, integrate-and-fire (IF) model is an important model for neurons. It has been considered for optogenetic systems in populations of coupled neurons \cite{nandi2017}, and (in our own previous work) for individual neurons subject to a distortion criterion on the output \cite{Noel2017d,Noel2018b}. \edit{In addition, optogenetic control strategies for \textit{ensembles} of neurons using this model were articulated in \cite{ching2013}.}{R2C7} The IF model considers neurons as capacitors, where the current is integrated over time to find the neuron's potential; once the potential exceeds a threshold, the neuron fires. IF is a first-order linear differential equation model, but its simplicity hides much of the complexity of real neurons. In particular, there are practical neuron behaviors that cannot be readily observed using the IF model; see \cite{Izhikevich2004}. Various other neuron models include linear models that address issues with IF, such as the leaky IF model, and nonlinear models, of which the Hodgkin-Huxley model \cite{Hodgkin1952} is likely the best known. \edit{Control strategies for neurons represented by a class of one-dimensional phase models were presented in \cite{nabi2012}, including a simplified version of the Hodgkin-Huxley model.}{R2C7} In this paper, we use a simplified (but realistic) nonlinear model known as the Izhikevich neuron model \cite{Izhikevich2003} (which we hereafter simply refer to as the \emph{Izhikevich model}).

\begin{figure}[!t]
    \centering
    \includegraphics[width=3.4in]{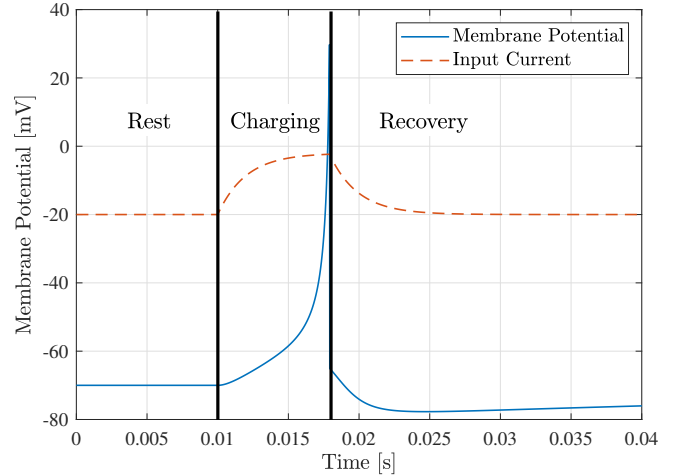}
    \caption{The neuron membrane behavior considered in this paper. The membrane potential versus time is plotted for the stimulation of one action potential. The membrane current is also shown on an arbitrary scale to indicate when the light source is turned on \edit{(i.e., when the current is increasing)}{}. The membrane potential starts at rest \edit{(``Rest'')}{}, and once the light source is turned on the membrane begins to charge \edit{(``Charging'')}{}. When the neuron is \edit{about}{} to fire, the light source is turned off. Finally, the membrane recovers \edit{(``Recovery'')}{} and returns to the resting potential.}
    \label{fig_single_spike}
\end{figure}

This paper considers how to control the optogenetic stimulation of neurons that follow the Izhikevich model, as summarized in Fig.~\ref{fig_single_spike}. 
The Izhikevich model is relatively simple to describe and simulate, but is biologically plausible because the range of neuron firing patterns that can be observed is consistent with all known types of cortical neurons, as demonstrated in \cite{Izhikevich2004} by tuning the model parameters. This is unlike other simple models, such as the IF model and its variants. The spiking patterns that can be generated using the Izhikevich model include the following: regular spiking (RS) neurons, in which spikes occur less frequently as stimulation is maintained; fast spiking (FS) neurons, where spiking at a high frequency can be maintained; low-threshold spiking (LTS) neurons, which are an intermediate between RS and FS; chattering (CH) neurons, in which spikes can occur in multiple bursts; and intrinsically bursting (IB), which can produce both regular spikes as well as irregular bursts. \edit{While the details of the Izhikevich model do not directly align with the biophysical mechanisms that underlie membrane potential dynamics (e.g., refractory periods are not clearly observed), its simplicity makes it amenable to the analysis that we undertake in this work. Models that include biophysical parameterization, e.g., the Hodgkin-Huxley family of models \cite{Hodgkin1952}, can be considered in future work.}{R2C2,R1C1}

The specific contributions of this work are as follows:
\begin{enumerate}
    \item We use curve fitting to estimate the illumination period required for an optogenetically-modified neuron to fire and recover to its resting potential, as a function of the Izhikevich model parameters. As in \cite{Noel2018}, our examples focus on RS neurons, but in this work we also give corresponding results for FS, LTS, and IB neurons\footnote{We note that ``chattering'' neurons do not align well with the methodology in this paper because, by design, they are prone to spiking multiple times \edit{even after the stimulating current is removed.}{}}. Furthermore, \edit{we adopt a more realistic optogenetic current model that is an approximation of results presented in \cite{Nikolic2009}, and we also fit the neuron behaviour to the peak}{R2C3} optogenetically-induced current. Our results show that our approach leads to accurate estimation of both the charging and recovery time, as measured by metrics including the mean squared error. This enables the generation of arbitrary spike sequences when there is sufficient time between consecutive spikes.
    \item We illustrate control of spike sequence generation by observing the distortion as a function of spike frequency. This expands the brief investigation of generating different spike frequencies in \cite{Noel2018}. We show how our numerical fits enable us to predict a lower bound on the achievable frequency without significant distortion. If additional distortion can be tolerated, then our results demonstrate that we can generate spikes at a target frequency that is up to \edit{about $50\,\%$ greater than}{} that predicted by our numerical method.
\end{enumerate}

The rest of this paper is organized as follows. Section~\ref{sec_models} describes the optogenetic and membrane potential models. We couple the two models in Section~\ref{section_sim}. We numerically fit the times for both charging and recovery, and observe the distortion as a function of a target firing frequency, in Section~\ref{section_fit}. We conclude in Section~\ref{sec_end}. 

\section{Physical Models}
\label{sec_models}


In this section, we briefly describe the two physical models that we integrate to describe the neuron stimulation and membrane potential. These are the optogenetic model for the external stimulation and the Izhikevich model for the membrane potential dynamics.

\subsection{Optogenetic System Model}

Neurons, like all animal cells, maintain an electric potential difference across their membranes. This {\em membrane potential} can be varied through the selective opening and closing of ion channels on the cell surface, allowing ions such as Na$^+$, Ca$^{2+}$, K$^+$, and Cl$^{2-}$ to flow across the membrane. Neurons have {\em voltage-gated} ion channels, which open in response to changes in the membrane potential. This sets up a positive feedback loop. For example, in {\em depolarization}, a stimulus causes Na$^+$ channels to open, thus raising the membrane potential, which causes more Na$^+$ channels to open, further raising the membrane potential, and so on. The resulting rapid change in membrane potential causes the neuron to ``fire''; see \cite{purves-book}.

Ion channels can also be {\em light-gated}, such that they open in response to light. A well-studied example of this is channelrhodopsin (ChR); see \cite{Nagel2002,Nagel2003,Deisseroth2017}. An optogenetically-modified neuron expresses light-gated channels in addition to voltage-gated channels. 
Thus, illuminating the neuron (for example with a laser) can initiate the firing of the neuron by triggering the initial flow of ions.

While the ion channel is open, the ion current passing through the channel is dependent on a number of environmental factors, including pH and ion concentration \cite{Nagel2003}. It can also depend on  the precise number and location of receptors on the surface of the neuron, which is usually unknown. Moreover, the dwell time in each channel state is a random variable. Works that model the states in detail include \cite{Nikolic2009}, \edit{and experimental results \cite{Nagel2002,Nagel2003} suggest that a neuron will experience a \emph{stable steady-state} current in response to a constant illumination intensity $\ell(t)$. In this work, we assume a constant illumination intensity that results in an approximation of the current model in \cite{Nikolic2009}. Specifically, if illumination starts at time $t=\ton$ and remain on, then we assume that current $I(t)$ passing through the membrane at time $t>\ton$ is of the form}{R2C3}
\begin{equation}
    I(t) = I(\ton)+ (\Imax-I(\ton))(1-e^{-(t-\ton)/\tauon}),
    \label{eqn_current}
\end{equation}
\edit{where $\tauon$ is an optogenetic time constant. Furthermore, if the illumination turns off at time $\toff$, then the current through the membrane decays according to}{R2C3}
\begin{equation}
    I(t) = I(\toff)e^{-(t-\toff)/\tauoff},
    \label{eqn_current_decay}
\end{equation}
\edit{where $\tauoff$ is another time constant. Using this current approximation, we find very good agreement with the membrane current dynamics shown in \cite[Fig.~9]{Nikolic2009} for short illumination times, i.e., on the order of less than 20\,ms, when we set $\tauon=\tauoff=2\,$ms (which we assume for the remainder of this work). By including $\ton$ and $\toff$, both (\ref{eqn_current}) and (\ref{eqn_current_decay}) also readily model the current if the illumination is repeatedly turned on and off.}{R2C3}


\subsection{Izhikevich Neuron Model}

The Izhikevich model uses a two-dimensional system of ordinary differential equations where the variables are the membrane potential $v(t)$ and the membrane recovery variable $u(t)$. $u(t)$, which accounts for the activation of potassium ionic current and the inactivation of sodium ionic currents, provides negative feedback to $v(t)$. The system of equations was obtained via fitting to natural spike initiation dynamics of cortical neurons and is as follows \cite[Eqs.~(1)--(3)]{Izhikevich2003}:
\begin{align}
\label{eqn_izhikevich_v}
    \frac{\mathrm{d}v(t)}{\mathrm{d}t} = &\, 0.04v^2 + 5v + 140 - u(t) + I(t), \\
\label{eqn_izhikevich_u}
    \frac{\mathrm{d}u(t)}{\mathrm{d}t} = &\, a(bv(t)-u(t)), \\
    \textrm{if } v(t) \geq &\, 30\,\textrm{mV, then}
    \left\{
	\begin{array}{l}
	v(t) \leftarrow c \\
\label{eqn_izhikevich_reset}
	u(t) \leftarrow u(t)+d,
	\end{array}
	\right. 
\end{align}
where (\ref{eqn_izhikevich_v}) and (\ref{eqn_izhikevich_u}) update the rates of change of $v(t)$ and $u(t)$, respectively, and (\ref{eqn_izhikevich_reset}) resets $u(t)$ and $v(t)$ after a spike occurs. Time and potential are measured in $\mathrm{ms}$ and $\mathrm{mV}$, respectively. $I(t)$ is the synaptic or input current through the ion channels in the dendrites and it is normalized. The parameters $a$, $b$, $c$, and $d$ are the fitting parameters and they can be tuned for different types of neurons; see \edit{\textit{nominal} values in Table~\ref{table_izhikevich_parameters}. We emphasise that parameter values are obtained by fitting to neuron membrane dynamics and thus vary for individual neurons.}{R3C2} $a$ sets the time scale of the decay of recovery variable $u(t)$ after a spike occurs. $b$ describes the sensitivity of $u(t)$ to subthreshold fluctuations of $v(t)$, and furthermore it can be used to define the membrane resting potential. $c$ is the reset potential for $v(t)$ after a spike occurs, and $d$ determines the reset of $u(t)$ after a spike occurs.

Results in \cite{Izhikevich2003,Izhikevich2004} demonstrate that the Izhikevich model can produce the behaviors of different types of cortical neurons by appropriately tuning the parameters $\{a,b,c,d\}$, even though the model itself is not analytically derived and so is not biophysically meaningful. Each type of neuron is associated with a characteristic firing pattern, where each firing pattern is a sequence of spikes. \edit{Nominal}{} model parameters for a selection of neuron types that are suitable for a broad range of neural behavior are listed in Table~\ref{table_izhikevich_parameters}.

\begin{table}[!t]
	\centering
	\caption{Selection of Nominal Parameter Values for the Izhikevich Neuron Model (from \cite{Izhikevich2003})}
	
	{\renewcommand{\arraystretch}{1.4}
		\begin{tabular}{l||c|c|c|c}
			\hline
			Neuron Type (Acronym) & $a$ & $b$ & $c$ & $d$ \\ \hline \hline
			Regular Spiking (RS) & 0.02 & 0.2 & -65 & 8 \\ \hline
			Fast Spiking (FS) & 0.1 & 0.2 & -65 & 2 \\ \hline
			Low-Threshold Spiking (LTS) & 0.02 & 0.25 & -65 & 2 \\ \hline
			Chattering (CH) & 0.02 & 0.2 & -50 & 2 \\ \hline
			Intrinsically Bursting (IB) & 0.02 & 0.2 & -55 & 4 \\ \hline
		\end{tabular}
	}
	\label{table_izhikevich_parameters}
\end{table}

\section{Simulating Neuron Spikes}
\label{section_sim}

In this section, we present the simulation of spikes in the Izhikevich model when stimulation is provided by the simple optogenetic model. \edit{First, we describe how we directly couple the two models. Next, we demonstrate the simulation of a sequence of spikes to motivate the selection of a suitable simulation time step. We also use these simulations to motivate our interest in studying individual spikes. Finally, we assess the impact of the model's initial conditions and derive the steady-state potentials of the Izhikevich model in the absence of an input current.}{R2C4}

\subsection{Coupling the Izhikevich Model with Optogenetics}

We take a direct approach to couple the two physical models for an individual neuron. We assume that the optogenetic stimulation is the membrane's \textit{only} external current source at the dendrites and it defines the input current $I(t)$ in (\ref{eqn_izhikevich_v}), \edit{such that membrane current is bounded within $0 \le I(t) \le \Imax$.}{R2C3} In practice, this is an approximation, since the Izhikevich model was initially developed for natural neurons, where input currents enter via the activation of neurotransmitter receptors at the dendrites. We assume that we can control where the light-gated channels are expressed in the membrane to imitate the conditions for the Izhikevich model. Otherwise, alternative means to describe the membrane dynamics would be required, which can be considered in future work. \edit{To use the simplified optogenetic model in (\ref{eqn_current}) and (\ref{eqn_current_decay}), we assume that we can turn the light source on and off as needed, and that it provides constant illumination $\ell(t)$ when the light source is on.}{R2C3} Thus, to simulate the complete system, we only need to initialize $\{u(t),v(t),I(t)\}$ and use (\ref{eqn_izhikevich_v})--(\ref{eqn_izhikevich_reset}) in a loop to update $u(t)$ and $v(t)$, where we update $I(t)$ or fire the neuron when required.

\subsection{\edit{Choice of Time Step}{}}
\begin{figure}[!t]
    \centering
    \includegraphics[width=3.4in]{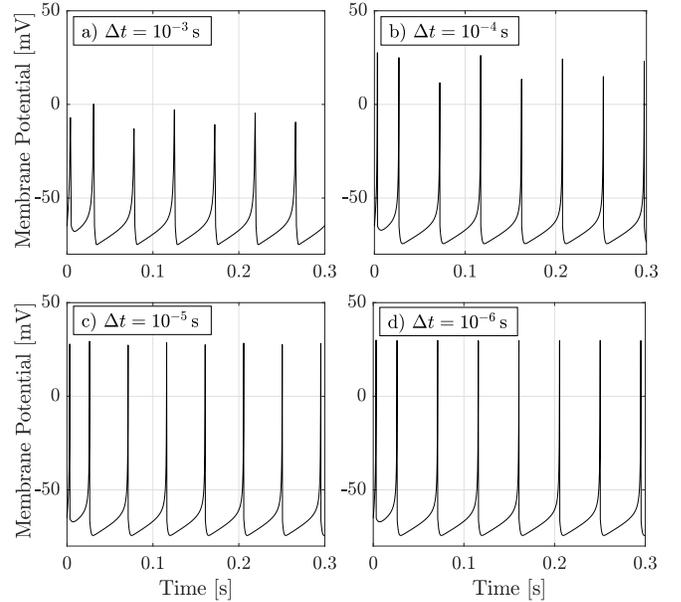}
    \caption{A sequence of neuron spikes for different values of time step $\dt$. The membrane is stimulated with a \textit{constant} current $I(t)=10$. The model parameters are $\{a,b,c,d\} = \{0.02,0.2,-65,8\}$ (i.e., \emph{Regular Spiking} in Table~\ref{table_izhikevich_parameters}).}
    \label{fig_time_step}
\end{figure}

We must choose a time step $\dt$ to set the resolution with which we evaluate (\ref{eqn_izhikevich_v})--(\ref{eqn_izhikevich_reset}). Specifically, $\dt$ is needed to update $u(t)$ and $v(t)$ from $\frac{\mathrm{d}u(t)}{\mathrm{d}t}$ and $\frac{\mathrm{d}v(t)}{\mathrm{d}t}$, respectively, i.e., we update $v(t)$ as
\begin{equation}
    v_\mathrm{new} = v_\mathrm{old} + \dt\frac{\mathrm{d}v(t)}{\mathrm{d}t}
    \label{eqn_v_update}
\end{equation}
and correspondingly update $u(t)$. In Fig.~\ref{fig_time_step}, we test different values of $\dt$ for a \emph{regular spiking} neuron by setting the (normalized) input current to a constant $I(t)=\Imax=10$ (practical values for plateau currents can be on the order of 100\,pA or more; see \cite{Nikolic2009,Grossman2011}). The default value of $\dt$ in \cite{Izhikevich2003,Izhikevich2004} is $\dt=10^{-3}\,\second$, but we see in Fig.~\ref{fig_time_step}a) that this results in an insufficient level of granularity for our analysis, i.e., $\frac{\mathrm{d}u(t)}{\mathrm{d}t}$ and $\frac{\mathrm{d}v(t)}{\mathrm{d}t}$ change too much over the scale of $\dt=10^{-3}\,\second$ to accurately update $v(t)$ in (\ref{eqn_v_update}). Thus, it appears that spikes are occurring before $v(t)$ reaches the \edit{spiking}{R3C3} potential of 30\,mV and furthermore that \edit{the spikes are occurring at random}{R3C3} potentials. This can be mitigated by decreasing $\dt$. However, decreasing $\dt$ also increases the computational resources required to simulate the neuron. The timing of the spikes is indistinguishable for $\dt=10^{-5}\,\second$ and $\dt=10^{-6}\,\second$, and \edit{the membrane potential for these cases always peaks at about 30\,mV for each spike,}{R3C3} but we use $\dt=10^{-6}\,\second$ in the remainder of this work to have sufficient resolution for the numerical fits. Unless otherwise stated, we also use $\Imax=6$.

\subsection{\edit{Generating Multiple Spikes}{}}
From Fig.~\ref{fig_time_step}, we also observe that the interspike intervals are not constant, \edit{and this is independent of the choice of $\dt$}{}. This behavior is expected for regular spiking neurons and other types of neurons as well. However, our objective is to fit expressions to describe a neuron's behavior and control when it fires. As an early work in this direction, we seek to ignore the effects of interspike interference, so we focus here on predicting the generation and recovery of individual spikes, as shown in Fig.~\ref{fig_single_spike}. We then use the results as a baseline for sequences of multiple spikes where the neuron is only stimulated while it is charging from rest. Repeated spiking patterns due to \emph{on-going} input current is a scenario for future work.

\subsection{Initial Conditions and the Steady State}

To maintain accuracy in our numerical analysis, we need to impose consistent conditions on the membrane. To generate a single spike, we will turn the \edit{illumination ``on'' until the neuron fires and then leave the illumination ``off''. In the absence of illumination, the membrane current will decrease to 0 and the membrane potential of the neuron should eventually converge to a}{R2C3} \emph{resting potential} (unless it is bistable or inhibition induced; see \cite{Izhikevich2004}). \edit{We can calculate the resting potential by}{R2C5} setting the left hand sides of (\ref{eqn_izhikevich_v}) and (\ref{eqn_izhikevich_u}) to $0$, \edit{setting the input current $I(t)$ to $0$, and then solving the two equations for $u(t)$ and $v(t)$}{R2C5}. From (\ref{eqn_izhikevich_u}) we can write $u(t) = bv(t)$, which we can substitute into (\ref{eqn_izhikevich_v}) and re-arrange for $v(t)$ to show that the two possible resting potentials are
\begin{equation}
\label{eqn_rest_potential}
    v_\mathrm{rest} = 12.5b - 62.5 \pm 12.5 \sqrt{b^2 -10b + 2.6}.
\end{equation}

The \edit{more negative}{} solution of (\ref{eqn_rest_potential}), $v_\mathrm{rest}^-$, is stable. The \edit{more positive}{} $v_\mathrm{rest}^+$ is unstable and is in fact \edit{the spike generation threshold. In other words, if the membrane potential is above $v_\mathrm{rest}^+$, then $v(t)$ will increase and the neuron will fire even if $I(t)=0$ (though firing within this model could still be avoided with a sufficiently large \emph{negative} current). Strictly speaking, we do not need to keep illuminating once the membrane potential reaches $v_\mathrm{rest}^+$, but we assume that it would be easier to detect when the \textit{peak} membrane potential is reached. Furthermore, automatic firing can still be accelerated by providing an input current, thus providing more precise control over firing times and a margin of error to avoid stopping illumination too early when generating multiple spikes.}{R3C3}

If the membrane potential is lower than $v_\mathrm{rest}^+$ and no input is applied, then the potential will converge to $v_\mathrm{rest}^-$. Throughout this work, we assume that the potential has converged once it remains within $\epsilon=0.5\%$ of $v_\mathrm{rest}^-$. We will see that this is a conservative estimate; in practice, we will not need to be so close to the resting potential before we can stimulate again without noticeable interspike interference.

\begin{figure}[!t]
    \centering
    \includegraphics[width=3.4in]{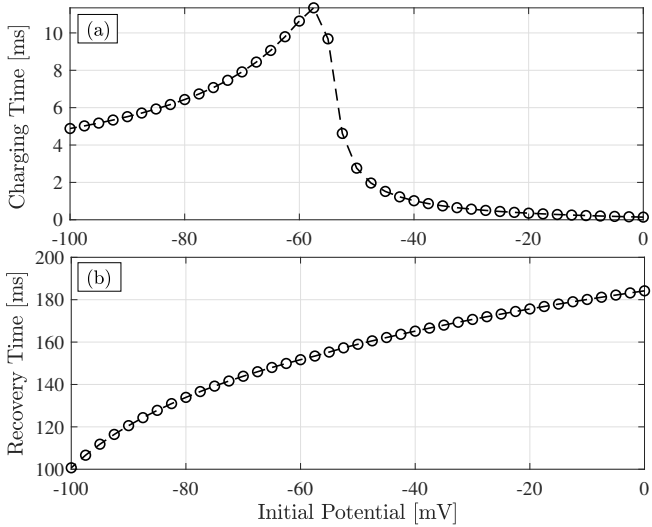}
    \caption{a) Charging time and b) recovery time for a single spike as a function of the initial membrane potential \edit{when illumination begins at time $t=0$ and $\Imax=6$. The illumination remains on}{} until the neuron fires. The model parameters are $\{a,b,c,d\} = \{0.02,0.2,-65,8\}$ (i.e., \emph{Regular Spiking} in Table~\ref{table_izhikevich_parameters}).}
    \label{fig_behavior_vs_initial}
\end{figure}

We refer to the time needed for the neuron to fire as the \emph{charging time} and the time to reach the stable resting potential as the \emph{recovery time}. We show in Fig.~\ref{fig_behavior_vs_initial}, where $v_\mathrm{rest}^-=-70\,\mathrm{mV}$, that \emph{both} of these times are sensitive to the initial membrane potential. To facilitate the application of this model to the generation of multiple spikes, \edit{we impose for the rest of this work}{R2C5} that the initial membrane potential is also the resting potential $v_\mathrm{rest}^-$, and that the \edit{recovery variable $u(0)$ is \textit{initially} $bv(0)$ (i.e., (4) is 0)}{R2C5}.

\section{Numerical Fitting Results}
\label{section_fit}

In this section, we assess whether we can predict the timing behavior, i.e., the charging and recovery times of the Izhikevich neuron model, based on knowledge of the model parameters. Specifically, we seek numerically-derived equations for a neuron's behavior as a function of $\{a,b,c,d,\Imax\}$. We are not predisposed towards any particular class of equations, but we seek results that are sufficiently accurate to use as a guide to control firing times and know how long to wait between firing times (i.e., for the membrane to return to the resting potential before we should start charging it again). Our assumptions limit the usefulness of very high precision; the optogenetic model \edit{is a simplifying approximation that produces results over short illumination periods that are consistent with \cite{Nikolic2009}}{R2C3}, the model parameters $\{a,b,c,d\}$ cannot be directly measured, and \edit{we assume that all of the models are deterministic, i.e., there are no}{} physical noise sources. However, the \edit{maximum current}{} $\Imax$ can be externally controlled to some extent by modifying the illumination intensity (though it \edit{may not}{} be constant in practice). We seek to gain intuition about controlling a neuron, and in particular we will estimate and measure the maximum firing frequency that can be achieved without interspike interference.

The remainder of this section is organized as follows. First, we measure the charging time and the recovery time as functions of the individual model parameters (including the \edit{maximum input}{} current $\Imax$), where the remaining model parameters are fixed.  This helps us decide which parameters to focus on in a joint model. For all types of neurons considered (RS, FS, LTS, and IB), the charging time is most sensitive to $b$ and $\Imax$ (we note that (\ref{eqn_izhikevich_v})--(\ref{eqn_izhikevich_reset}) show that charging time is independent of $c$ and $d$), and the recovery time is most sensitive to $a$ and $d$. Next, we measure the charging time as a function of both $b$ and $\Imax$ and the recovery time as a function of both $a$ and $d$. All fitting functions are found via nonlinear least squares in MATLAB using the \texttt{fit} function with default tolerances. Finally, we consider the stimulation of multiple spikes, where we predict the interference-free firing frequency and measure the deviations from target firing times as a function of the target firing frequency.

\subsection{Fitting to a Single Spike}

As we are primarily interested in the charging time and recovery time for each neuron type, we use curve fitting to develop accurate models for these properties under various parameter values. This is a challenging task given the five-dimensional parameter space. We first consider fits to individual parameters, keeping other parameters at a ``typical'' value for a particular neuron type, and then consider fits to multiple parameters. We give a detailed explanation and analysis of our method using a regular spiking (RS) neuron as an example; results for the other types of neurons are summarised in the corresponding tables.

We measure the accuracy of the fitting functions with three methods. $R^2$ measures the proportion of the variance in the behavior that is predictable from the model parameters, where $R^2 \in [0,1]$. The root mean square error (RMSE) measures the standard deviation of the behavior from that predicted by the fitting functions. The maximum error (Max Error) is simply the absolute value of the largest deviation from the fitting function over the parameter range or ranges considered. Consider that we are fitting to a total of $N$ parameter value combinations (where we vary one or more of the parameters $\{a,b,c,d,\Imax\}$). We then suppose that $y_n$ is the charging time (in ms) for the $n$th combination of the model parameters, $\hat{y}_n$ is the corresponding estimated charging time due to some fitting function, and $\overline{y}$ is the average charging time over all $N$ model parameter combinations. A similar description can be made for recovery time. Then, $R^2$ for the charging time is measured as
\begin{equation}
    R^2 = 1 - \frac{\sum_{n=1}^N (y_n-\hat{y}_n)^2}{\sum_{n=1}^N (y_n-\overline{y})^2},
\end{equation}
the RMSE is measured in ms as
\begin{equation}
    \mathrm{RMSE} = \sqrt{\frac{\sum_{n=1}^N (y_n-\hat{y}_n)^2}{N}},
    \label{eqn_rmse}
\end{equation}
and the maximum error in ms is
\begin{equation}
    \mathrm{Max~Error} = \underset{n}{\max} |y_n-\hat{y}_n|.
\end{equation}

To fit behavior to the individual parameters, we consider polynomial functions up to degree 4 (i.e., from linear to quartic, beyond which minimal improvement was observed), exponential functions with either 1 or 2 terms, and power functions of the form $y = nx^m + p$. These fitting functions were the most relevant in MATLAB's Curve Fitting Toolbox. To fit the behavior to the individual model parameters, we vary one parameter while holding the remaining parameters constant. The chosen range of each parameter is in consideration of the types of neurons listed in Table~\ref{table_izhikevich_parameters}. Using the RS neuron as an example, our default parameter values are $\{a,b,c,d\} = \{0.02,0.2,-65,8\}$, which is consistent with RS, \edit{and our default maximum current is $\Imax=6$}{}. If we vary one of the parameters $\{a,b,c,d,\Imax\}$, then the remainder are fixed at the default value. The range of each varied parameter, a selection of fitted equations for their behavior (chosen for quality and space), and the accuracy of each fit are summarized in Table~\ref{table_single_charging} for charging time and Table~\ref{table_single_recovery} for recovery time (see the Appendix for additional functions that fit the behavior of the RS neuron to the \edit{maximum current}{} $\Imax$). \edit{Our considered parameter ranges varied for different neuron types in order to guarantee that the neuron would fire \textit{and} would not fire more than once.}{} We note that, as we might expect from Table~\ref{table_izhikevich_parameters}, some of the fits for different neuron types are identical because there are common parameter values \edit{and ranges}{}. This is particularly the case for charging time because it is only a function of two of the Izhikevich model parameters. For example, since RS, FS, and IB neurons all have the same nominal value of $b$ \edit{and the same range for $a$,}{} they also have the same fitting function of $a$ for charging time.

\begin{table*}[!t]
	\centering
	\caption{Fitting charging behavior to a single parameter. Default parameter values are from Table~\ref{table_izhikevich_parameters} and $\Imax=6$. Increment over the parameter range is 0.005 for $a$, 0.005 for $b$, and 0.5 for $\Imax$.}
\resizebox{\textwidth}{!}{%
	{\renewcommand{\arraystretch}{1.4}
		\begin{tabular}{l|c||c|c|c|c|c}
			\hline
			\thead{Neuron \\ Type} & \thead{Parameter \\ Range} & Fit & Fitting Function & $R^2$ & \thead{RMSE \\ ~[ms]} & \thead{Max\\ Error [ms]} \\ \hline \hline
			\multirow{3}{*}{RS} & $a \in [0.02,0.1]$ & poly1 & $4.003a + 7.834 $ & $0.9999 $ & $3.72\powten{-4} $ & $9.61\powten{-4} $ \\ \cline{2-7}
			& $b \in [0.2,0.25]$ & power1 &
			$0.2922b^{-2.050}  $ & $0.9997 $ & $1.545\powten{-2} $ & $3.769\powten{-2} $ \\ \cline{2-7}
			& $\Imax \in [4,12]$ & power2 &
			$69.28\Imax^{-1.512} + 3.317 $ & $0.9995 $ & $4.584\powten{-2} $ & $8.790\powten{-2} $ \\ \hline
			\multirow{3}{*}{FS} & $a \in [0.02,0.1]$ & poly1 & $4.003a +7.834 $ & $0.9999 $ & $3.72\powten{-4} $ & $9.61\powten{-4} $ \\ \cline{2-7}
			& $b \in [0.2,0.25]$ & power1 & $0.2505b^{-2.169} $ & $0.9998 $ & $1.246\powten{-2} $ & $2.632\powten{-2} $ \\ \cline{2-7}
			& $\Imax \in [4,12]$ & power2 & $150.6\Imax^{-1.993}+4.002 $ & $0.9980 $ & $10.67\powten{-2} $ & $0.2001 $ \\ \hline
			\multirow{3}{*}{LTS} & $a \in [0.02,0.1]$ & poly1 & $0.8328a+4.958 $ & $0.9998 $ & $2.94\powten{-4} $ & $5.1\powten{-4} $ \\ \cline{2-7}
			& $b \in [0.2,0.25]$ & power1 & $0.2922^{-2.050} $ & $0.9997 $ & $1.549\powten{-2} $ & $3.769\powten{-2} $ \\ \cline{2-7}
			& $\Imax \in [2,12]$ & power1 & $11.37\Imax^{-0.4602} $ & $0.9999 $ & $7.259\powten{-3} $ & $1.507\powten{-2} $ \\ \hline
			\multirow{3}{*}{IB} & $a \in [0.02,0.1]$ & poly1 & $4.003a+7.834 $ & $0.9999 $ & $3.72\powten{-4} $ & $9.61\powten{-4} $ \\ \cline{2-7}
			& $b \in [0.18,0.21]$ & power1 & $0.2256I^{-2.212} $ & $0.9989 $ & $3.175\powten{-2} $ & $4.816\powten{-2} $ \\ \cline{2-7}
			& $\Imax \in [4,12]$ & power1 & $46.60\Imax^{-0.9916} $ & $0.9951 $ & $1.068\powten{-1} $ & $0.1376 $ \\ \hline
		\end{tabular}
	}
}
	\label{table_single_charging}
\end{table*}

\begin{table*}[!t]
	\centering
	\caption{Fitting recovery behavior to a single parameter. Default parameter values are from Table~\ref{table_izhikevich_parameters} and $\Imax=6$. Increment over the parameter range is 0.005 for $a$, 0.005 for $b$, 1 for $c$, 0.5 for $d$, and 0.5 for $\Imax$.}
\resizebox{\textwidth}{!}{%
	{\renewcommand{\arraystretch}{1.4}
		\begin{tabular}{l|c||c|c|c|c|c}
			\hline
			\thead{Neuron \\ Type} & \thead{Parameter \\ Range} & Fit & Fitting Function & $R^2$ & \thead{RMSE \\ ~[ms]} & \thead{Max\\ Error [ms]} \\ \hline \hline
			\multirow{5}{*}{RS} & $a \in [0.02,0.1]$ & power1 & $3.371a^{-0.9587}$ & $1.000 $ & $0.2696 $ & $0.5037 $ \\ \cline{2-7}
			& $b \in [0.2,0.25]$ & poly3 &
			$ -1.831\powten{5}b^3 + 1.161\powten{5}b^2 - 2.451\powten{4}b +1869$ & $0.9976 $ & $ 0.1646$ & $0.2718 $ \\ \cline{2-7}
			& $c \in [-65,-50]$ & exp2 &
			$147.0e^{3.341 \times 10^{-4}c}+9.6847\powten{13}e^{0.6390c} $ & $0.9971 $ & $2.799\times 10^{-2} $ & $8.074\times 10^{-2} $ \\ \cline{2-7}
			& $d \in [2,10]$ & power1 &
			$78.30b^{0.2945} $ & $0.9994 $ & $0.3756 $ & $0.7580 $ \\ \cline{2-7}
			& $\Imax \in [4,12]$ & power2 &
			$23.14\Imax^{-2.4628} + 143.6$ & $0.9992 $ & $5.703\times10^{-3} $ & $1.352\times 10^{-2} $ \\ \hline
			\multirow{5}{*}{FS} & $a \in [0.02,0.1]$ & power1 & $3.367a^{-0.8522} $ & $0.9996 $ & $0.4221 $ & $0.8628 $ \\ \cline{2-7}
			& $b \in [0.2,0.25]$ & poly3 & $-1.233\powten{4}b^3 +7614b^2-1548b+128.2 $ & $0.9998 $ & $2.544\powten{-3} $ & $3.650\powten{-3} $ \\ \cline{2-7}
			& $c \in [-65,-55]$ & poly4 & \thead{$1.559\powten{-3}c^4+0.3819c^3$ \\ $+35.07c^2+1431c+2.193\powten{4} $} & $0.9973 $ & $5.336\powten{-2} $ & $0.1212 $ \\ \cline{2-7}
			& $d \in [2,10]$ & power1 & $21.54d^{0.1736} $ & $0.9966 $ & $0.1170 $ & $0.2609 $ \\ \cline{2-7}
			& $\Imax \in [4,12]$ & exp2 & $8.572e^{-0.4449\Imax}+22.75e^{8.625\powten{-3}\Imax} $ & $0.9961 $ & $1.418\powten{-2} $ & $2.463\powten{-2} $ \\ \hline
			\multirow{5}{*}{LTS} & $a \in [0.02,0.1]$ & power1 & $3.298a^{-0.8500} $ & $0.9986 $ & $0.7094 $ & $1.338 $ \\ \cline{2-7}
			& $b \in [0.2,0.25]$ & exp2 & $-8.864\powten{-10}e^{92.44b} +71.04e^{1.1474b}$ & $0.9999 $ & $1.613\powten{-2} $ & $3.052\powten{-2} $ \\ \cline{2-7}
			& $c \in [-70,-61]$ & poly3 & $1.030\powten{-2}c^3+2.079c^2+140.3c+3252$ & $0.9958 $ & $8.735\powten{-2} $ & $0.1786 $ \\ \cline{2-7}
			& $d \in [2,10]$ & power1 & $77.51d^{0.2603} $ & $1.000 $ & $7.412\powten{-2} $ & $0.1997 $ \\ \cline{2-7}
			& $\Imax \in [4,10]$ & poly3 & $4.151\powten{-2}\Imax^3-0.7104\Imax^2+4.274\Imax+84.02 $ & $0.9928 $ & $0.1189 $ & $0.2278 $ \\ \hline
			\multirow{5}{*}{IB} & $a \in [0.02,0.1]$ & power1 & $3.478a^{-0.9040} $ & $0.9997 $ & $0.4039 $ & $0.7936 $ \\ \cline{2-7}
			& $b \in [0.18,0.21]$ & exp1 & $97.54e^{1.048b} $ & $0.9989 $ & $4.104\powten{-2} $ & $7.065\powten{-2} $ \\ \cline{2-7}
			& $c \in [-65,-55]$ & poly3 & $2.667\powten{-3}c^3+0.4995c^2+31.27c+773.0$ & $0.9981 $ & $2.675\powten{-2} $ & $5.120\powten{-2} $ \\ \cline{2-7}
			& $d \in [3.5,8]$ & power1 & $83.66d^{0.2623} $ & $0.9998 $ & $0.1395 $ & $0.3388 $ \\ \cline{2-7}
			& $\Imax \in [4,6.5]$ & poly2 & $0.2143\Imax^2-2.190\Imax+125.7 $ & $0.9986 $ & $5.295\powten{-3} $ & $7.629\powten{-3} $ \\ \hline
		\end{tabular}
	}
}
	\label{table_single_recovery}
\end{table*}

\begin{figure}[!t]
    \centering
    \includegraphics[width=3.5in]{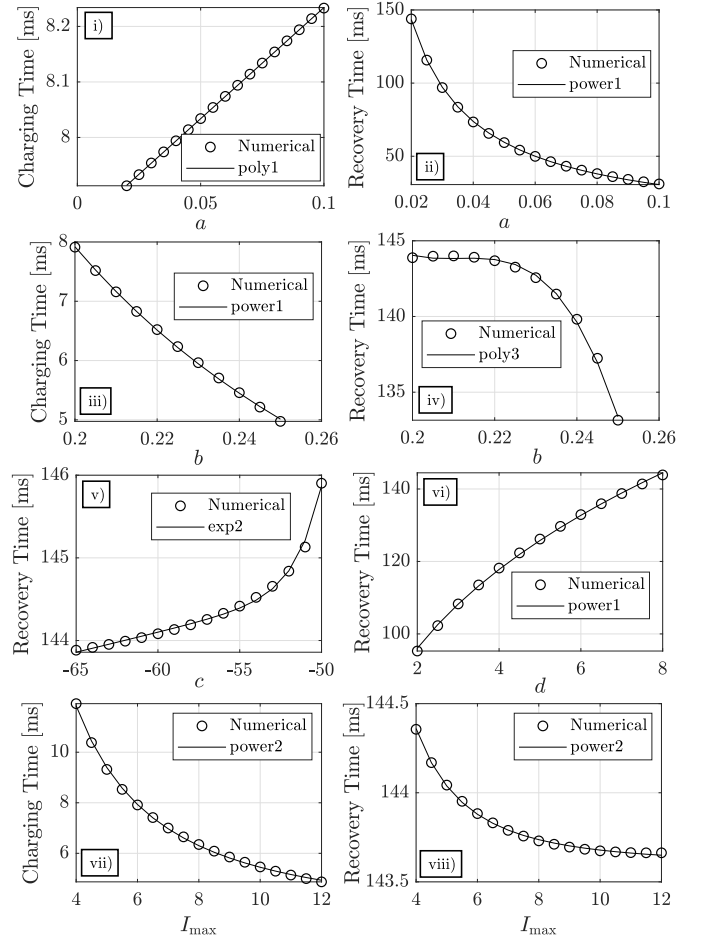}
    \caption{Charging and recovery time behavior for an RS neuron as a function of individual model parameters, each shown with one representative numerical fit. Both charging and recovery times are found as functions of $\{a,b,\Imax\}$. Recovery times are also found as functions of $\{c,d\}$. The nominal parameter values are $\{a,b,c,d,\Imax\} = \{0.02,0.2,-65,8,6\}$. \edit{The illumination remains on until the neuron fires.}{}}
    \label{fig_rs_single}
\end{figure}

In Fig.~\ref{fig_rs_single}, we plot the charging and recovery times for a single spike of a nominal RS neuron while varying one individual model parameter. A representative numerical fit accompanies each plot, and is generally chosen to be the simplest fit that results in $R^2 > 0.995$. The results are generally consistent with the other types of neurons that we consider, and are also consistent with what we would expect given (\ref{eqn_izhikevich_v})-(\ref{eqn_izhikevich_reset}).

The charging time in Fig.~\ref{fig_rs_single} depends on $\{a,b,\Imax\}$. While the charging time is nearly independent of $a$, it noticeably decreases with increasing $b$ or $\Imax$. The recovery time depends on all of the model parameters, but is nearly independent of $c$ and $\Imax$. It is not surprising for the \edit{maximum magnitude of the current to not have a significant impact on the the recovery time, since the illumination is always turned off during recovery and the current decays.}{} However, it might be surprising that parameter $c$, which via (\ref{eqn_izhikevich_reset}) dictates the reset potential after the neuron fires, has a negligible impact on the time to recover. This is due to the exponential recovery behavior. The recovery time is most sensitive to $a$ and $d$.

Perhaps with the exception of the \edit{maximum}{} current $\Imax$, because it is an external and controllable parameter, fitting to multiple model parameters is preferable. So, based on the single-parameter fitting for an RS neuron in Fig.~\ref{fig_rs_single}, we consider two-parameter fits for an RS neuron. In particular, we fit to the charging time by varying $b$ and $\Imax$, and we fit to the recovery time by varying $a$ and $d$. We consider polynomial surfaces up to degree 4, where for simplicity both parameters always have the same degree. We hold the remaining model parameters constant according to the nominal parameter values in Table~\ref{table_izhikevich_parameters}. A fitted surface for each type of neuron and the accuracy of each fit are summarized in Table~\ref{table_multi_charging} for charging time and Table~\ref{table_multi_recovery} for recovery time (see the Appendix for additional functions that fit the charging time of the RS neuron to the current $\Imax$ and parameter $b$).


\begin{table*}[!t]
	\centering
	\caption{Fitting charging behavior to $b$ and $\Imax$. Default parameter values are from Table~\ref{table_izhikevich_parameters}. Increment over the parameter range is 0.005 for $b$ and 0.5 for $\Imax$.}
\resizebox{\textwidth}{!}{%
	{\renewcommand{\arraystretch}{1.4}
		\begin{tabular}{l|c||c|c|c|c}
			\hline
			\thead{Neuron \\ Type} & \thead{Parameter \\ Range} & Fitting Function & $R^2$ & \thead{RMSE \\ ~[ms]} & \thead{Max\\ Error [ms]} \\ \hline \hline
			RS & \thead{$b \in [0.2,0.25]$ \\ $\Imax \in [4,12]$} & \thead{$187.3 -17.58I-1553b +0.6887I^2 + 90.70\Imax b +4686b^2 $ \\ $-9.523\powten{-3}\Imax^3-1.765\Imax^2b-118.5\Imax b^2-5115b^3 $} & $0.9962 $ & $9.117\powten{-2} $ & $0.6975 $ \\ \hline
			{FS} & \thead{$b \in [0.2,0.25]$ \\ $\Imax \in [4,12]$} & \thead{$269.0 -24.42\Imax-2353b +0.9272\Imax^2 + 131.7\Imax b +7360b^2$ \\ $ -1.265\powten{-2}\Imax^3-2.437\Imax^2b-181.2\Imax b^2-8159b^3 $} & $0.9913 $ & $0.1546 $ & $1.329 $ \\ \hline
			{LTS} & \thead{$b \in [0.2,0.25]$ \\ $\Imax \in [4,10]$} & \thead{$228.1 -23.99\Imax-1922b +1.033\Imax^2 + 127.3\Imax b +5820b^2$ \\ $ -1.577\powten{-2}\Imax^3-2.719\Imax^2b-171.5\Imax b^2-6306b^3 $} & $0.9975 $ & $7.345\powten{-2} $ & $0.5075 $ \\ \hline
			IB & \thead{$b \in [0.18,0.21]$ \\ $\Imax \in [4,6.5]$} & \thead{$3.900\powten{4}-6257\Imax-6.137\powten{5}b+477.6\Imax^2+6.808\powten{4}\Imax b $ \\ $+3.702\powten{6}b^2 -18.93I^3-3233\Imax^2b-2.533\powten{5}\Imax b^2-1.011\powten{7}b^3 $ \\ $+0.3230\Imax^4+59.29I^3b+5665\Imax^2b^2+3.205\powten{5}\Imax b^3+1.052\powten{7}b^4 $} & $0.9933 $ & $0.2793 $ & $0.8903 $ \\ \hline
			
		\end{tabular}
	}
}
	\label{table_multi_charging}
\end{table*}

\begin{table*}[!t]
	\centering
	\caption{Fitting recovery behavior to $a \in [0.02,0.1]$ and $d \in [2,10]$ (except for IB neurons, where $d \in [3.5,8]$). Default parameter values are from Table~\ref{table_izhikevich_parameters} and $\Imax=6$. Increment over the parameter range is 0.005 for $a$ and 0.5 for $d$.}
\resizebox{\textwidth}{!}{%
	{\renewcommand{\arraystretch}{1.4}
		\begin{tabular}{l||c|c|c|c}
			\hline
			\thead{Neuron \\ Type} & Fitting Function & $R^2$ & \thead{RMSE \\ ~[ms]} & \thead{Max\\ Error [ms]} \\ \hline \hline
			RS, FS & \thead{$153.5 + 19.16d -5491a -1.011d^2 -301.7ad +7.784\powten{4}a^2$ \\  $+2.552\powten{-2}d^3 +6.646ad^2 +1389a^2d -3.658\powten{5}a^3 $} & $0.9950 $ & $1.924 $ & $5.673 $ \\ \hline
			LTS & \thead{$154.8 +14.51d -5358a -0.6033d^2 -248.3ad +7.490\powten{4}a^2 $ \\ $+9.958\powten{-3}d^3 +5.148ad^2 +1154a^2d -3.484\powten{5}a^3 $} & $0.9949 $ & $1.801 $ & $5.260 $ \\ \hline
			IB & \thead{$180.5 +11.60d -6048a -0.1827d^2 -229.3ad +8.375\powten{4}a^2 $ \\ $-6.411\powten{-3}d^3 +3.422ad^2 +1126a^2d -3.892\powten{5}a^3 $} & $0.9953 $ & $1.952 $ & $5.416 $ \\ \hline
			
		\end{tabular}
	}
}
	\label{table_multi_recovery}
\end{table*}

\begin{figure}[!t]
    \centering
    \includegraphics[width=3.5in]{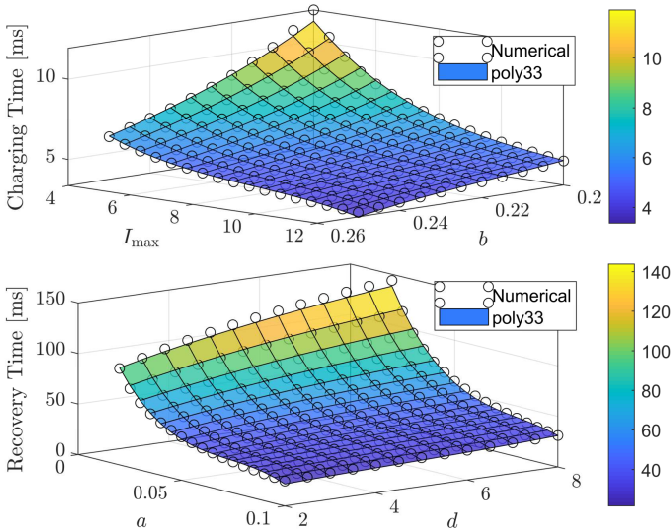}
    \caption{Charging and recovery time behavior for an RS neuron as a function of two model parameters, each shown with one representative numerical fitting surface. The charging time (Top) is found as a function of $b$ and $\Imax$. The recovery time (Bottom) is found as a function $a$ and $d$. The nominal parameter values are $\{a,b,c,d,\Imax\} = \{0.02,0.2,-65,8,6\}$. \edit{The illumination remains on until the neuron fires.}{}}
    \label{fig_rs_multi}
\end{figure}

In Fig.~\ref{fig_rs_multi}, we plot the charging time as a function of $b$ and $\Imax$ and the recovery time as a function of $a$ and $d$ for a nominal RS neuron. We include the third order polynomial surface fit for \edit{both the charging time and the}{} recovery time. Both surface fits agree with the numerical data, as indicated in Tables~\ref{table_multi_charging} and \ref{table_multi_recovery}. We can see that the charging time is sensitive to both $b$ and $\Imax$ for the entire range of parameter values considered, whereas the recovery time is relatively more sensitive to $a$ than to $d$.

\begin{figure}[!t]
    \centering
    \includegraphics[width=3.5in]{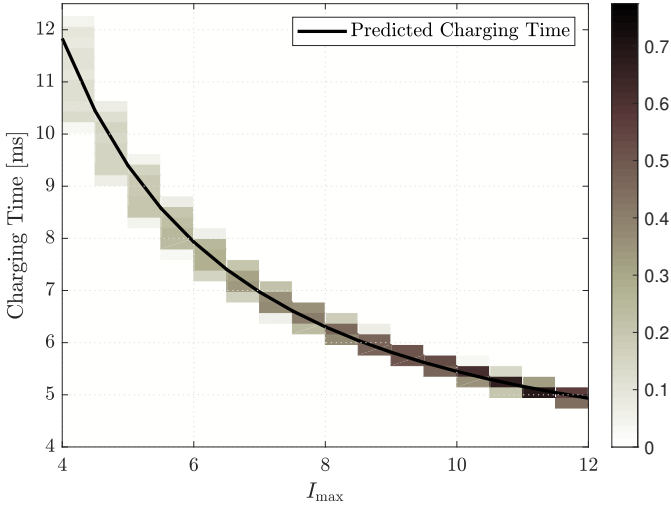}
    \caption{Distribution of charging times (from rest) for a set of $10^3$ non-identical RS neurons as a function of the \edit{maximum}{} input current. Parameters $a$ and $b$ are uniformly chosen over ranges $a \in [0.02,0.036]$, $b \in [0.2,0.21]$, i.e., over 20\% of the value ranges in Table~\ref{table_izhikevich_parameters}. The distributions are compared with the expected charging time given the nominal $a$ and $b$ values for an RS neuron.}
    \label{fig_rs_firing_distribution}
\end{figure}

One might question how reliably we can depend on the particular model parameter values if the Izhikevich model itself was obtained via numerical fitting to experimental data. Since the charging time is generally much faster than the recovery time, we measure the sensitivity of the charging time to \emph{random} model parameters $a$ and $b$ in Fig.~\ref{fig_rs_firing_distribution}, where we predict the charging time as a function of the \edit{maximum}{} stimulation current $\Imax$ and we assume that the $a$ and $b$ parameters are the nominal values for an RS neuron, i.e., $\{a,b\} = \{0.02, 0.2\}$. For each considered value of $\Imax$, we generate $10^3$ realizations of $a$ and $b$ parameters that are uniformly distributed over the ranges $a \in [0.02,0.036]$, $b \in [0.2,0.21]$, calculate the charging time from rest for each realization by solving (\ref{eqn_izhikevich_v})--(\ref{eqn_izhikevich_reset}), and then plot the distribution of the charging times. Fig.~\ref{fig_rs_firing_distribution} shows that the actual charging times deviate from the predicted value \edit{by less than 10\% for most of the range of maximum}{} currents $\Imax \in [4,12]$.

\subsection{Applying Fits to Spike Trains}

Thus far, we have focused on the generation of individual spikes in isolation. However, we can apply our results to the generation of spike trains. In particular, we can assess how well we can generate a spike train at a target frequency, where the period is the sum of the charging and recovery times. We turn the \edit{illumination}{} on for the expected time to charge the neuron, leave the \edit{illumination}{} off for the neuron to recover, and then charge the neuron again. We are interested in measuring the \textit{deviation} in spike times from the target frequency when we provide insufficient time for the neuron to fully recover. Similar to our work with the IF neuron model in \cite{Noel2017d, Noel2018b}, we can use (\ref{eqn_rmse}) to calculate the RMSE associated with a spike train of $N$ spikes, but where $y_n$ is the $n$th target firing time (according to a specified firing frequency), and $\hat{y}_n$ is the corresponding observed firing time.

\begin{figure}[!t]
    \centering
    \includegraphics[width=3.5in]{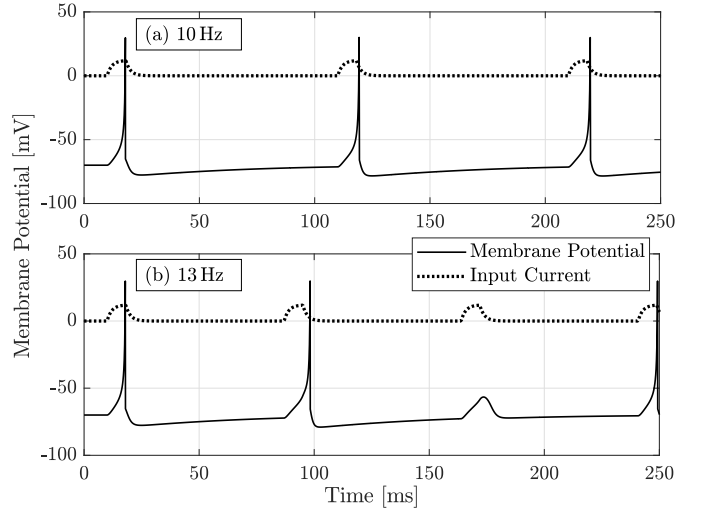}
    \caption{Membrane potential versus time for an RS neuron ($\{a,b,c,d\} = \{0.02,0.2,-65,8\}$) that is stimulated with a \edit{maximum current $\Imax=6$}{} to fire at specified frequencies. The input current is drawn on an arbitrary scale to show when it turns on and off.}
    \label{fig_rs_vs_time_freq}
\end{figure}

\begin{figure}[!t]
    \centering
    \includegraphics[width=3.5in]{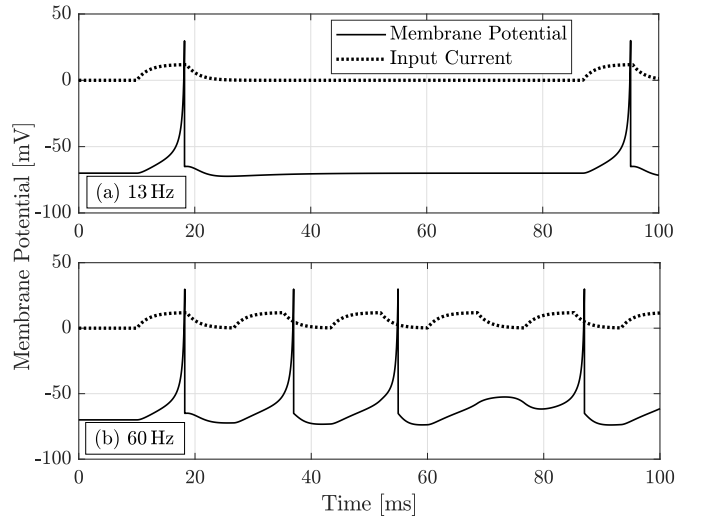}
    \caption{Membrane potential versus time for an FS neuron ($\{a,b,c,d\} = \{0.1,0.2,-65,2\}$) that is stimulated with a \edit{maximum current $\Imax=6$}{} to fire at specified frequencies. The input current is drawn on an arbitrary scale to show when it turns on and off.}
    \label{fig_fs_vs_time_freq}
\end{figure}

First, we observe deviations visually. Using Tables~\ref{table_multi_charging} and \ref{table_multi_recovery}, we expect an RS neuron with nominal model parameters and \edit{maximum}{} current $\Imax=6$ to take $8.13$\,ms to charge and $138.2$\,ms to recover. Thus the interference-free firing frequency is approximately 6.8\,Hz. In Fig.~\ref{fig_rs_vs_time_freq}, we observe the input current and membrane potential of an RS neuron versus time as we try to generate spikes at 10\,Hz and 13\,Hz, where in each case we turn on the stimulating current for $8.13$\,ms. At 10\,Hz, we observe that the spikes can still be generated but that deviations from the target firing time are visually apparent with the third spike (\edit{since we expect the neuron to fire}{} as soon as the current is turned off). At 13\,Hz, there is a more visible deviation with the second spike and then the third spike is missed entirely. We can achieve faster controlled spiking with a different neuron type. In Fig.~\ref{fig_fs_vs_time_freq}, we observe the input current \edit{and}{} membrane potential of an FS neuron, which we can calculate has an interference-free firing frequency of approximately 33.1\,Hz. A spike train at 13\,Hz can be generated without a problem, but a 60\,Hz spike train misses spikes.

\begin{figure}[!t]
    \centering
    \includegraphics[width=3.5in]{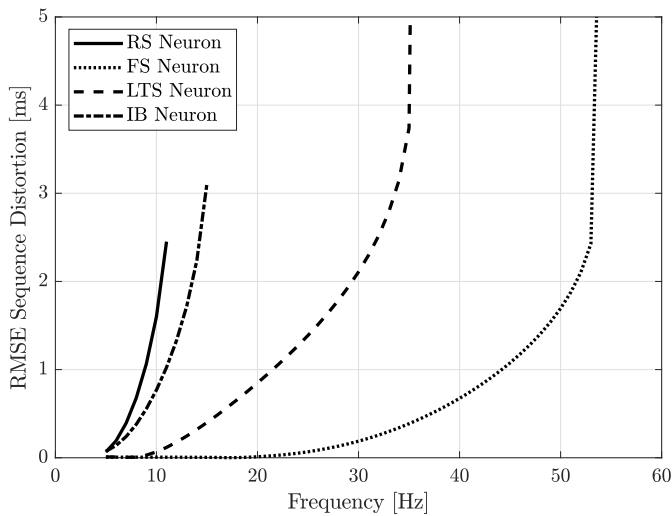}
    \caption{Root mean square distortion versus firing frequency for the RS, FS, LTS, and IB neuron types. The model parameter values are those that are nominal for each type of neuron (as listed in Table~\ref{table_izhikevich_parameters}), the \edit{maximum current is $\Imax=6$, and the neuron is illuminated when we expect to be charging it.}{} The charging time was found using the fitting functions for $\Imax$ in Table~\ref{table_single_charging}. The distortion is measured relative to a target sequence of the same frequency whose firing times are synchronous with the first expected firing time. The second through eleventh firing times are considered to calculate the RMSE.}
    \label{fig_distortion_vs_freq}
\end{figure}

To provide more detailed insight into the generation of spike trains at different frequencies, we measure the RMSE for sequences of 10 spikes (\emph{after} the distortion-free first spike) as a function of the target firing frequency for RS, FS, LTS, and IB neurons in Fig.~\ref{fig_distortion_vs_freq} where we set the \edit{maximum}{} current $\Imax=6$ and found the charging time from $\Imax$ in Table~\ref{table_single_charging}. For each type of neuron, the distortion jumps to infinity when we miss a spike. RS neurons are the least accommodating of rapid stimulation, followed by IB neurons, LTS neurons, and then FS neurons. Generally, for each type of neuron, the maximum possible frequency without missing spikes is \edit{less than}{} twice that predicted by the interference-free charging and recovery times, e.g., 11\,Hz for the RS neuron and 53\,Hz for the FS neuron. \edit{Thus, the interference-free estimates provide a ``rule-of-thumb'' to predict achievable firing times in a spike train.}{}

\begin{figure}[!t]
    \centering
    \includegraphics[width=3.5in]{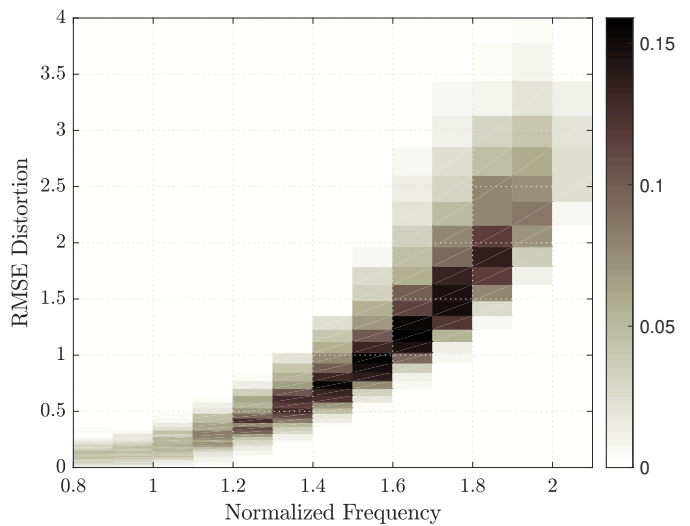}
    \caption{Distribution of distortion for a set of $10^3$ non-identical FS neurons as a function of normalized frequency. The second through eleventh firing times are considered. Each neuron has target model parameter values that are uniformly chosen over ranges $a \in [0.084,0.1]$, $b \in [0.2,0.21]$, $c \in [-65,-62]$, $d \in [2,3.2]$, i.e., over 20\% of the value ranges in Table~\ref{table_izhikevich_parameters}. Actual model parameter values are normally distributed about the target parameter values with variances that are 1\% of the chosen ranges. Frequencies are normalized to that predicted by the corresponding charging and recovery times in Tables~\ref{table_multi_charging} and \ref{table_multi_recovery}, respectively. The \edit{maximum current is $\Imax=6$}{}.}
    \label{fig_distortion_distribution}
\end{figure}

Finally, we measure the distribution of sequence distortions for FS neurons, where we set $\Imax=6$ and generate $10^3$ realizations of target FS model parameters over the ranges $a \in [0.084,0.1]$, $b \in [0.2,0.21]$, $c \in [-65,-62]$, and $d \in [2,3.2]$. For each realization of target model parameters, we generate actual parameter values that are normally distributed about the target values and with variances that are 1\% of the chosen ranges. We use Tables~\ref{table_multi_charging} and \ref{table_multi_recovery} to determine the target charging and recovery times, and then in Fig.~\ref{fig_distortion_distribution} measure the distribution of RMSE distortion as a function of the normalized frequency. The frequencies are normalized to the frequency predicted by the charging and recovery times in Tables~\ref{table_multi_charging} and \ref{table_multi_recovery}. Even though we are simulating neurons with model parameters that do not match those used to predict the charging and recovery times, the results in Fig.~\ref{fig_distortion_distribution} are still consistent with those in Fig.~\ref{fig_distortion_vs_freq}, such that FS neurons can be stimulated with RMSE distortion usually below 1\,ms if the firing frequency is no more than \edit{$50\,\%$ greater than}{} that predicted by the charging and recovery times. This demonstrates the robustness of our methodology to control the firing of individual neurons.

\section{Conclusion}
\label{sec_end}

In this paper, we have considered the use of an optogenetic stimulation model to control the timing of individual neuron spikes. We used the Izhikevich neuron model for the membrane potential dynamics and fitted the neuron charging and recovery times to functions of the model's parameters and the input current. We have demonstrated that simple functions can help predict lower bounds on the highest firing frequency that can be achieved in \emph{regular spiking}, \emph{fast spiking}, \emph{low-threshold spiking}, and \emph{intrinsically bursting} neurons with minimal interspike interference. We have also measured deviations due to imperfect knowledge of the neuron model parameters. \edit{Future work can consider mismatch between neuron model parameters and experimentally-observed membrane potential dynamics,}{R2C6} develop a new model for membrane potential dynamics to align with where light-gated channels are expressed and opened, and study information-theoretic measures for the information that can be embedded in externally-stimulated spike trains.

\appendix

In Table~\ref{table_rs_current}, we list additional equations found for fitting the charging and recovery times of the \edit{nominal}{} RS neuron to the \edit{maximum}{} input current $\Imax$. In Table~\ref{table_rs_current_joint}, we list additional equations found for fitting the charging times of the \edit{nominal}{} RS neuron to the \edit{maximum}{} input current $\Imax$ and model parameter $b$.

\begin{table*}[!t]
	\centering
	\caption{\edit{Fitting behavior of RS neuron to maximum current $\Imax \in [4,12]$}{}. Model parameter values are $\{a,b,c,d\} = \{0.02,0.2,-65,8\}$. Increment of $\Imax$ over the parameter range is 0.5.}
\resizebox{\textwidth}{!}{%
	{\renewcommand{\arraystretch}{1.4}
		\begin{tabular}{l||c|c|c|c|c}
			\hline
			Behavior & Fit & Fitting Function & $R^2$ & RMSE [ms] & Max Error [ms] \\ \hline \hline
			\multirow{8}{*}{Charging}
			& poly1 & $-0.7515\Imax+13.00 $ & $0.8704 $ & $0.7102 $ & $1.9256 $ \\ \cline{2-6}
			& poly2 &	$0.1229\Imax^2-2.718\Imax+20.13 $ & $0.9810 $ & $0.2717 $ & $0.6966 $ \\ \cline{2-6}
			& poly3 &	$-2.209\powten{-2}\Imax^3+0.6532\Imax^2-6.723\Imax+29.54 $ & $0.9971 $ & $0.1058 $ & $0.2326 $ \\ \cline{2-6}
			& poly4 &	$4.133\powten{-3}\Imax^4-0.1544\Imax^3+2.177\Imax^2-14.18\Imax+42.55 $ & $0.9996 $ & $4.123\powten{-2} $ & $8.338\powten{-2} $ \\ \cline{2-6}
			& power1 &	$35.70\Imax^{-0.8232} $ & $0.9891 $ & $0.2064 $ & $0.5178 $ \\ \cline{2-6}
			& power2 &	$69.28\Imax^{-1.512}+3.317 $ & $0.9995 $ & $4.584\powten{-2} $ & $8.790\powten{-2} $ \\ \cline{2-6}
			& exp1 &	$16.97e^{-0.1161\Imax} $ & $0.9370 $ & $0.4954 $ & $1.261 $ \\ \cline{2-6}
			& exp2 &	$69.34e^{-0.7032\Imax} +9.784e^{-5.901\powten{-2}\Imax}$ & $0.9998 $ & $2.455\powten{-2} $ & $5.011\powten{-2} $ \\ \hline
			\multirow{8}{*}{Recovery}
			& poly1 & $-7.007\powten{-2}\Imax+144.4 $ & $0.7614 $ & $9.607\powten{-2} $ & $0.2563 $ \\ \cline{2-6}
			& poly2 &	$1.679\powten{-2}\Imax^2 -0.3387\Imax+145.4$ & $0.9690 $ & $3.461\powten{-2} $ & $8.838\powten{-2} $ \\ \cline{2-6}
			& poly3 &	$-2.825\powten{-3}\Imax^3+8.458\powten{-2}\Imax^2-0.8506I+146.6 $ & $0.9955 $ & $1.324\powten{-2} $ & $2.907\powten{-2} $ \\ \cline{2-6}
			& poly4 &	$5.180\powten{-4}\Imax^4-1.940\powten{-2}\Imax^3+0.2756I^2-1.785I+148.2 $ & $0.9993 $ & $5.111\powten{-3} $ & $1.041\powten{-2} $ \\ \cline{2-6}
			& power1 &	$145.0I^{-3.878\powten{-3}} $ & $0.8724 $ & $7.027\powten{-2} $ & $0.1783 $ \\ \cline{2-6}
			& power2 &	$23.14I^{-2.463}+143.6 $ & $0.9992 $ & $5.703\powten{-3} $ & $1.352\powten{-2} $ \\ \cline{2-6}
			& exp1 &	$144.4e^{-4.875\powten{-4}\Imax} $ & $0.7618 $ & $9.599\powten{-2} $ & $0.2559 $ \\ \cline{2-6}
			& exp2 &	$7.081e^{-0.5910I}+143.7e^{-2.868\powten{-5}\Imax} $ & $0.9995 $ & $4.280\powten{-3} $ & $8.278\powten{-3} $ \\ \hline
		\end{tabular}
	}
	}
	\label{table_rs_current}
\end{table*}

\begin{table*}[!t]
	\centering
	\caption{\edit{Fitting charging behavior of RS neurons to $b \in [0.2,0.25]$ and $\Imax \in [4,12]$}{}. Default parameter values are from Table~\ref{table_izhikevich_parameters}. Increment over the parameter range is 0.005 for $b$ and 0.5 for $\Imax$.}
\resizebox{\textwidth}{!}{%
	{\renewcommand{\arraystretch}{1.4}
		\begin{tabular}{c|c|c|c|c}
			\hline
			\thead{Fit} & Fitting Function & $R^2$ & \thead{RMSE \\ ~[ms]} & \thead{Max\\ Error [ms]} \\ \hline \hline
			poly11 & $20.23 -0.4692I -48.34b $ & $0.8687 $ & $0.5366 $ & $3.187 $ \\ \hline
			poly22 & $54.68 -3.526I-249.5b +6.301\powten{-2}\Imax^2 + 9.105Ib +285.1b^2 $ & $0.9788 $ & $0.2156 $ & $1.540 $ \\ \hline
			poly33 & \thead{$187.3 -17.58I-1553b +0.6887I^2 + 90.70Ib +4686b^2 $ \\ $-9.523\powten{-3}\Imax^3-1.765I^2b-118.5Ib^2-5115^3 $} & $0.9962 $ & $9.117\powten{-2} $ & $0.6975 $ \\ \hline
			poly44 & \thead{$591.3 -82.86I -6603b +4.757I^2 +685.3Ib +2.838\powten{4}b^2 -0.1351I^3 -24.82I^2b -1962Ib^2$ \\ $ -5.414\powten{4}b^3 +1.525\powten{-3}\Imax^4 +0.3412I^3b +33.03I^2b^2 +1948Ib^3 +3.716\powten{4}b^4 $} & $0.9993 $ & $3.962\powten{-2} $ & $0.2983 $ \\ \hline
			
		\end{tabular}
	}
}
	\label{table_rs_current_joint}
\end{table*}

\bibliography{2018_journal_izhikevich}

\end{document}